\documentclass[prd, floatfix,
 reprint,
superscriptaddress,
groupedaddress,
unsortedaddress,
nofootinbib,
 amsmath,amssymb,
 aps,
]{revtex4-2}
\usepackage{amsmath}
\usepackage{amssymb}

\usepackage{graphicx}
\usepackage[dvipsnames]{xcolor}
\definecolor{light gray}{gray}{0.8}
\usepackage{pgfplots}
\usepackage{tikz}
\usetikzlibrary{decorations.markings} 

\usepackage{placeins}
\usepackage{dcolumn}
\usepackage{bm}
\definecolor{prdlink}{RGB}{0,0,128}\usepackage[colorlinks=true,linkcolor=prdlink,citecolor=prdlink,urlcolor=prdlink]{hyperref}
\usepackage{orcidlink}

\pgfplotsset{compat=1.18}
\begin{document}

\preprint{APS/123-QED}

\title{Time delay in matched exterior and interior Kottler solutions}

\author{Mourad Guenouche\,\orcidlink{0000-0002-1211-4940}}
\email{guenouche\_mourad@umc.edu.dz}\email{guenouche\_mourad@univ-khenchela.dz}
  
\affiliation{Laboratoire de Physique Théorique, \href{https://ror.org/017wv6808}{Université Frères Mentouri-Constantine 1}, BP 325 route de Ain El Bey, 25017 Constantine, Algeria \\
and Département des Sciences de la Matière, \href{https://ror.org/02yyskm09}{Université Abbès Laghrour-Khenchela}, BP 1252, El Houria, Route de Constantine, 40004 Khenchela, Algeria}

\date{\today}

\begin{abstract}
We extend the study of gravitational lensing to photons traversing the lens mass, modeled as a uniform-density fluid described by the interior Kottler (Schwarzschild–de Sitter) solution smoothly matched to the exterior Kottler region. An analytic expression for the time delay is derived, allowing the interior contribution to be explicitly isolated relative to the vacuum Kottler case. This correction is found to systematically enhance the total time delay, an effect corroborated by numerical evaluations for astrophysically relevant lenses at both galaxy and cluster scales. These results underscore the importance of accounting for the interior lens structure in accurate modeling of strong lensing time delays.
\end{abstract}

\maketitle
\section{Introduction}
Gravitational lensing time delays provide a powerful probe of both cosmic expansion and the internal structure of astrophysical lenses \cite{Refsdal}. Standard analyses usually treat the lens as a point mass embedded in the Schwarzschild–de Sitter (Kottler) spacetime, neglecting the interior contributions of the mass distribution \cite{Schu2}. While this approximation is sufficient for compact objects, it becomes inadequate for extended systems such as galaxies and clusters, where photons may traverse the lens interior. In such cases, the time delay acquires corrections beyond the exterior vacuum contribution.

In this work we model the lens as a uniform-density fluid sphere described by the interior Kottler solution, smoothly matched at the boundary to the exterior Kottler metric. This construction extends the classical Schwarzschild interior–exterior matching to include a cosmological constant, thereby providing a consistent framework for light propagation through extended matter distributions in $\Lambda$CDM cosmology. By integrating the null geodesic equations across both regions, we derive analytic expressions for the bending angle and the travel-time difference between multiple photon trajectories.

The paper is organized as follows. In Sec.~\ref{II} we review the matched interior–exterior Kottler construction and the associated geodesic equations. In Sec.~\ref{III} we derive the analytic expression for the time delay and isolate the interior contribution relative to the vacuum case. Numerical results for representative galaxy- and cluster-scale lenses are presented in Sec.~\ref{IV}. Finally, Sec.~\ref{V} summarizes our findings and concludes with perspectives.

\section{\label{II}Matching interior and exterior Kottler metrics}
The exterior and interior SdS metrics can be expressed together by the line element
\begin{equation}
{\rm d}s^{2}=f(r){\rm d}t^{2}-g(r)^{-1}{\rm d}r^{2}-r^{2}\left( {\rm d}%
\theta ^{2}+\sin ^{2}\theta {\rm d}\varphi ^{2}\right),  \label{SdSint}
\end{equation}
where the metric functions $f(r)$ and $g(r)$ are related by
\begin{widetext}
\begin{equation}
f(r)= \left\{ 
\begin{array}{lll}
g(r), & g(r)=1-\delta(r)-\lambda(r)^{2}, & r\geq r_{\rm B} \\ 
\left[ \dfrac{3}{2}\dfrac{\delta _{\rm B}}{\delta_{\rm B}+\lambda_{\rm B}^2}\sqrt{g_{\rm B}}
-\left( \dfrac{3}{2}\dfrac{\delta _{\rm B}}{\delta_{\rm B}+\lambda_{\rm B}^2}-1\right) \sqrt{g(r)}
\right] ^{2}, & g(r)=1-\left(\delta_{\rm B}+\lambda_{\rm B}^2\right)\dfrac{r^{2}}{r_{\rm B}^{2}}, & 
r\leq r_{{\rm B}}
\end{array}
\right.,
\end{equation}
\end{widetext}
using auxiliary functions defined by $\delta (r)=2GM/r$, $\lambda (r)=\sqrt{\Lambda /3}r$. Here, $2GM$ is the Schwarzschild radius in the geometrized units $c=1$, and $M$ is the total mass. The subscript B denotes the evaluation of these functions at the boundary radius of the sphere fluid $r=r_{{\rm B%
}}$, where the exterior ($r\geq r_{{\rm B}}$) and the interior ($r\leq r_{%
{\rm B}}$) SdS metrics are smoothly matched, provided $f_{\rm B}=g_{\rm B}=1-\delta_{\rm B}-\lambda_{\rm B}^2$, with $g_{\rm B}=g(r_{\rm B})$, $\delta_{\rm B}=2GM/r_{\rm B}$, and $\lambda_{\rm B}=\sqrt{\Lambda/3}r_{\rm B}$. 

Assuming a uniform fluid density, satisfying $4\pi r_{{\rm B}}^{3}\rho =3M$, the pressure profile within the mass distribution is isotropic and can be expressed as
\begin{equation}
p(r)=\rho \frac{\sqrt{g(r)}-\sqrt{g_{\rm B}}}{3\left( 1-\lambda _{\rm B
}^{2}/\delta _{\rm B}\right) ^{-1}\sqrt{g_{\rm B}}-\sqrt{g(r)}}\quad \left( r\leq r_{\rm B}\right),
\end{equation}
which vanishes at the boundary surface $r=r_{\rm B}$, consistent with the matching to the exterior vacuum solution, but increases monotonically inward and diverges at the center $r=0$ if the compactness parameter $\delta _{\rm B}$ attains the critical value
\begin{equation}
\delta _{\rm B}=\frac{4}{9}\left( 1+\sqrt{1-\frac{9}{4}\lambda _{\rm B}^{2}}\right).
\end{equation}
Such a singular situation is clearly prevented under physically relevant configurations characterized  by extremely low compactness,
\begin{equation}
\delta _{\rm B}\ll   \frac{8}{9}-\frac{1}{2}\lambda _{\rm B}^{2}\quad \left( \lambda _{\rm B}^{2}\ll1\right),\label{bound}
\end{equation}
ensuring finite pressure throughout the fluid sphere. This condition holds for stellar or fluid structures with radii much larger than their Schwarzschild radii, which are ubiquitous in the Universe, except for compact bodies such as neutron stars and black holes. Note that the bound in (\ref{bound}) generalizes the Buchdahl limit \cite{Buchdahl}, $\delta _{\rm B}\leq8/9$, for a stable Schwarzschild perfect fluid, to include the repulsive effect of $\Lambda$. However, the assumption of uniform density implies that sound waves propagate instantaneously throughout the mass distribution, thereby violating causality. Although the inclusion of $\Lambda$ modifies the pressure profile, it does not resolve this issue unless a more realistic non-uniform density profile were employed \cite{Gabbanelli}.
\section{\label{III}Integrating null geodesic equations}
\begin{figure}[ht]
\centering
\begin{tikzpicture}
 
\fill[gray!62] (0,0) circle [radius=1.2cm]; 
\node at (1.3,-0.6) {L};
 
\draw[->, thin ] (-4.2,0) -- (4.2,0) node[pos=0.25, below] {$r_{\rm E}$} node[pos=1, above]{$x$};

\draw[dashed](0,0) -- (4.2,-0.5)  node[pos=0.5, below] {$r_{\rm S}$};

\fill[darkgray!85] (0.85,0.85) circle [radius=0.05cm];
\draw[dashed, ->, thin](0,0) -- (0.85,0.85) node[pos=0.6, right]{$r_{\rm B}$};

\fill[darkgray!85] (-0.734,0.952) circle [radius=0.05cm];
\draw[dashed, ->, thin](0,0) -- (-0.734,0.952) node[pos=0.6, left]{$r_{\rm B}$};

\fill[darkgray!85] (0.62,-1.03) circle [radius=0.05cm];
\draw[dashed, ->, thin](0,0) -- (0.62,-1.03) node[pos=0.6, right]{$r_{\rm B}$};

\fill[darkgray!85] (-0.734,-0.952) circle [radius=0.05cm];
\draw[dashed, ->, thin](0,0) -- (-0.734,-0.952) node[pos=0.6, left]{$r_{\rm B}$};

\draw[dashed, thin](0,0) -- (-0.09,1.05) node[pos=0.7, right]{$r_{0}$};
\draw[dashed, thin](0,0) -- (-0.09,-1.05) node[pos=0.75, right]{$r_{0}^\prime$};

\draw[thin, black] (2.8,0) arc (0:-19.5:1cm) node[pos=0.5, right]{$-\varphi_\text{S}$};

\draw[thin, black] (-3.35,0) arc (0:20.35:1cm) node[pos=0.5, right]{$\alpha$};
\draw[thin, black] (-3.2,0) arc (0:-20.35:1cm) node[pos=0.5, right]{$\alpha^\prime$};

\coordinate (A) at (4.2,-0.5);
\coordinate (C) at (-4.2,0);
\draw[color=red, postaction={ decorate, decoration={ markings, mark=at position 0.3 with {\arrow{>}}, mark=at position 0.5 with {\arrow{>}}, mark=at position 0.7with {\arrow{>}}}}] (A) to [bend left =-29] (C);

\draw[color=red, postaction={ decorate, decoration={ markings, mark=at position 0.3 with {\arrow{>}}, mark=at position 0.5 with {\arrow{>}}, mark=at position 0.7 with {\arrow{>}}}}] (A) to [bend right =-18] (C);

\fill[red] (4.2,-0.5) circle [radius=0.08cm] node[above, text=black]{S};
\fill[blue] (-4.2,0) circle [radius=0.08cm] node[above, text=black]{E};
\fill[black] (0,0) circle [radius=0.03cm];

\end{tikzpicture} 
\caption{Two light rays emitted by a source S, bent outside and inside the SdS fluid lens L, and received at the blue Earth E.}
\label{fig}
\end{figure}
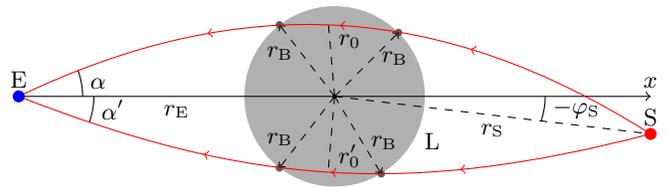

We consider two photons experiencing deflection as they pass near a spherically symmetric mass $M$ (the lens) Fig.~\ref{fig}, and compute the corresponding difference in their travel times---the time delay. Both photons are emitted by a source S, at emission times $t_{\rm S}$ and $t_{\rm S}^\prime$, pass through the mass distribution of radius $r_{\rm B}$, and are detected simultaneously on Earth E at $t_{\rm E}=t_{\rm E}^\prime=0$. Hence, $-t_{\rm S}$ and $-t_{\rm S}^\prime$ represent the positive travel times of the two photons, and the time delay reduces to $\Delta t=t_{\rm S}-t_{\rm S}^\prime$. We denote by $\alpha_{\rm E}$ and $\alpha_{\rm E}^\prime$ the reception angles of the two photons with respect to the lens-Earth axis. Let $r_0$ ($\leq r_{\rm B}$) and $r_{0}^\prime$ ($\leq r_{\rm B}$) be the respective distances of closest approach of the two trajectories, hereafter referred to as the pericenter distances. We restrict ourselves to cases where the pericenter is much smaller than radial distances $r_{\rm E}$ and $r_{\rm S}$ where located the Earth and the source, i.e., $r_{0}/r_{\rm E}\ll 1$ and $r_{0}/r_{\rm S}\ll 1$, conditions typically realized 
in light-deflection experiments. We choose the polar axis of the spherical coordinate system such that 
Earth and source are characterized by $\varphi=\varphi_{\rm E}=\pi$ and $\varphi=-\varphi_{\rm S}$. 

Owing to spherical symmetry, photon trajectories are restricted to a plane, which can, without loss of generality, be taken as the equatorial plane $\theta =\pi /2$. Consequently, only the $t$ and $\varphi $ components of the null  geodesic equations are needed. These take the same form in both the exterior and interior SdS metrics, namely,
\begin{equation}
\frac{\ddot{t}}{\dot{t}}+\frac{\dot{f}(r)}{f(r)}=0,\quad \frac{\ddot{\varphi}%
}{\dot{\varphi}}+2\frac{\dot{r}}{r}=0,  \label{geod1}
\end{equation}
where $\dot{\phantom{x}}={\rm d}/{\rm d}\tau$, with $\tau $ an affine parameter distinct from the proper time $s$. Each of the equations above is associated with a conserved quantity, $E=\dot{t}f(r)$ and $J=\dot{\varphi}r^{2}$, interpreted respectively as the conserved energy and angular momentum per unit mass. Inserting these into the normalization condition $\dot{s} =\varepsilon $ $(=0$ for null geodesics) yields the effective radial equation
\begin{equation}
f(r)\left( \frac{\dot{r}^{2}}{g(r)}+\frac{J^{2}}{r^{2}}\right) =E^{2}.
\label{radialeqt}
\end{equation}
Dividing $\dot{\varphi}$ $(=J/r^{2})$ and $\dot{t}$ $(=E/f(r))$ by $\dot{r}$ (\ref{radialeqt}) leads to differential equations governing the angular (azimuthal) $\varphi $ and temporal $t$ variations with respect to the radial coordinate $r$,
\begin{align}
{\rm d}\varphi (r)&=\pm \frac{{\rm d}r}{u(r)},\quad u(r)=r\sqrt{
g(r)\left( \frac{f_{0}r^{2}}{r_{0}^{2}f(r)}-1\right) },  \label{defeqt} \\
{\rm d}t(r)&=\pm \frac{{\rm d}r}{v(r)},\quad v(r)=\sqrt{f(r)g(r)\left( 1-
\frac{r_{0}^{2}f(r)}{f_{0}r^{2}}\right) },  \label{TDeqt}
\end{align}
where the pericenter $r_{0}$ is defined to satisfy the equation $u(r_{0})=0$, implying $J/E=\pm r_{0}/\sqrt{f_{0}}$, often interpreted (in absence of $\Lambda$) as the impact parameter, with $f_{0}=f(r_{0})$ for $r_{0}\leq r_{\rm B}$. This can be expressed in terms of observable quantities as $J/E=\pm r_{{\rm %
E}}\sin \alpha /\sqrt{f_{\rm E}}$ with $f_{\rm E}=f(r_{\rm E})$, by invoking the well-known final exterior SdS-type conditions that specify the 4-velocity of the photon upon its arrival at Earth, $(\dot{t}_{\rm E},\dot{r}_{\rm E},\dot{\theta}_{\rm E},\dot{\varphi}_{\rm E})=(1,\pm f_{\rm E}\cos \alpha ,0,\pm \sqrt{f_{\rm E}}\sin \alpha /r_{\rm E})$. These conditions are derived using the definition, $\tan\alpha =r_{\rm E}\sqrt{f_{\rm E}}{\rm d}\varphi(r_{\rm E})/{\rm d}r$. Then, a specific relation follows for calculating $r_{0}$,
\begin{equation}
\frac{r_{0}}{r_{\rm E}\sin \alpha }=\sqrt{\frac{f_{0}}{f_{\rm E}}}.
\label{r0rel}
\end{equation}
Note that the plus and minus signs in Eq. (\ref{defeqt}) reflect, respectively, outgoing and ingoing geodesics, relative to the lens position, and that the positive or negative $J$ correspond to increasing (upper trajectory) or decreasing $\varphi$ (lower trajectory) along the path.

For the upper photon, $r
$ is decreasing over $[ r_{\rm S},r_{\rm B}] $ and $[
r_{\rm B},r_{0}] $, while it is increasing over $[ r_{0},r_{%
\rm B}] $ and $[ r_{\rm B},r_{\rm E}]$. Then, integrating Eqs.~(\ref{defeqt}) and (\ref{TDeqt}) allows one to write the inclination angle $-\varphi _{\rm S}$ (azimuthal angle of the source) and the photon emission time $-t_{\rm S}$ at the source as
\begin{align}
-\varphi _{{\rm S}} &= \left( \int_{r_{\rm B}}^{r_{\rm E}}+\int_{r_{
\rm B}}^{r_{\rm S}}\right) \frac{{\rm d}r}{u_{{\rm ext}}(r)}
+2\int_{r_{0}}^{r_{\rm B}}\frac{{\rm d}r}{u_{{\rm int}}(r)}-\pi,
\label{def}\\
-t_{\rm S} &=\left( \int_{r_{{\rm B}}}^{r_{{\rm E}}}+\int_{r_{{\rm B}
}}^{r_{{\rm S}}}\right) \frac{{\rm d}r}{v_{{\rm ext}}(r)}
+2\int_{r_{0}}^{r_{{\rm B}}}\frac{{\rm d}r}{v_{{\rm int}}(r)},
\label{Ttravel}
\end{align}
where the upper photon is considered to be received on Earth at the azimuthal angle $\varphi _{\rm E}=\varphi (r_{\rm E})=\pi $ and the coordinate time $t_{\rm E}=t(r_{\rm E})=0$, with $u_{{\rm ext}}(r)$, $u_{{\rm int}}(r)$, $v_{{\rm ext}}(r)$ and $v_{{\rm int}}(r)$ the corresponding expressions of $u(r)$ (\ref{defeqt}) and $v(r)$ (\ref{TDeqt}) for the exterior and interior SdS metrics.

In the exterior SdS solution, the quantities $u(t)$ and $v(t)$ in Eqs.~(\ref{defeqt}) and (\ref{TDeqt}) are generically evaluated in terms of elementary functions, via a perturbation expansion in the small parameter $\delta _{0}=\delta(r_0)=2GM/r_{0}\ll 1$. This is valid for most non-compact astrophysical bodies whose sizes greatly exceed their Schwarzschild radii. Importantly, we suggest extending this for photons crossing into the interior SdS spacetime, in the sense that approximating in terms of $\delta_{\rm B}$ amounts to the same thing, provided $2GM\ll r_0\leq r_{\rm B}$.

\subsection{Bending of light}
Here, we aim to derive an analytical expression for $\varphi_{\rm S}$, which characterizes the deflection of light by the gravitational lens. Approximating $u_{\rm ext}(r)$ and $u_{\rm int}(r)$ to first order in $\delta_0$, and putting $x=r_0/r$, one gets
\begin{align}
u_{{\rm ext}}(x)^{-1} &\simeq \frac{x^{2}}{r_{0}\sqrt{1-x^{2}}}\left( 1+\frac{
\delta _{0}}{2}\frac{h_{0}-x^{3}}{1-x^{2}}\right),\label{uext} \\
u_{{\rm int}}(x)^{-1} & \simeq \frac{x^{2}}{r_{0}\sqrt{1-x^{2}}}\left[1+3\frac{\delta _{\rm B}}{2} \frac{\sqrt{1-\lambda _{\rm B}^{2}}\sqrt{1-\lambda _{0}^{2}}}{\lambda _{\rm B}^{2}\left( 1-x^{2}\right) }\right.\nonumber\\
&\left.\times\left( \dfrac{\sqrt{1-\lambda _{0}^{2}}x}{\sqrt{x^{2}-\lambda _{0}^{2}}}  -1\right) \right],\label{uint}
\end{align}
where, for convenience, a positive definite function $h(r)$ is further introduced,
\begin{equation}
h(r)=\dfrac{r^3}{r_{\rm B}^3}\left( 3\frac{1-\sqrt{1-\lambda _{\rm B}^{2}}\sqrt{1-\lambda (r)^{2}}}{\lambda (r)^{2}}-2 \right)\quad \left( r\leq r_{\rm B}\right),
\end{equation}
with $0<h(r)\leq 1$, which is involved in the approximate expression of the interior SdS metric function $f(r)$ to first order in $\delta_0$,
\begin{equation}
f(r)\simeq 1-\delta (r)h(r)-\lambda (r)^{2}\quad \left( r\leq r_{\rm B}\right),
\end{equation}
provided,
\begin{align}
f_{0}&\simeq 1-\delta _{0}h_{0}-\lambda _{0}^{2}{},\\
h_{0}&=h(r_0)=\dfrac{r_{0}^3}{r_{{\rm B}}^3}\left( 3\frac{1-\sqrt{1-\lambda _{{\rm B}}^{2}}\sqrt{1-\lambda
_{0}^{2}}}{\lambda _0^{2}}-2\right).
\end{align}
One can infer that $u_{{\rm ext}}(x)$ and $u_{{\rm int}}(x)$ coincide at the boundary $x_{\rm B}=x(r_{\rm B})=r_0/r_{\rm B}$ and that $h(r_{\rm B})=1$, ensuring continuity. Also, the perturbative term in $u_{{\rm int}}$ is indeed first order in $\delta_{{\rm B}}$, as can be checked through the identity,
\begin{equation}
\frac{\sqrt{1-\lambda _{0}^{2}{}}x  -\sqrt{x^{2}-\lambda _{0}^{2}}}{\lambda _{{\rm B}}^{2}}=\frac{r_{0}^{2}}{r_{{\rm B}}^{2}}\frac{1-x^{2}}{\sqrt{1-\lambda _{0}^{2}}x+\sqrt{x^{2}-\lambda_{0}^{2}}}.\label{identity}
\end{equation}
Making use of standard integrals
\begin{align}
&\int \dfrac{{\rm d}x}{\sqrt{1-x^{2}}}=\arcsin x,\label{integ1}\\
&\int \dfrac{{\rm d}x}{\left( 1-x^{2}\right) \sqrt{1-x^{2}}}= 
\frac{x}{\sqrt{1-x^{2}}},  \label{integ2} \\
&\int \frac{x^{3}{\rm d}x}{\left( 1-x^{2}\right) \sqrt{1-x^{2}}}=
\sqrt{1-x^{2}}+\frac{1}{\sqrt{1-x^{2}}},  \label{integ3} 
\end{align}
the first integral of the exterior region in (\ref{def}) is evaluated to
\begin{align}
\int_{r_{{\rm B}}}^{r_{{\rm E}}}\frac{{\rm d}r}{u_{{\rm ext}}(r)}
&\simeq-\arcsin x_{{\rm E}}+\arcsin x_{{\rm B}} \nonumber\\
&+\frac{\delta _{0}}{2}\left( \sqrt{1-x_{{\rm E}}^{2}}+\frac{1-h_{0}x_{
{\rm E}}}{\sqrt{1-x_{{\rm E}}^{2}}}\right.  \nonumber\\
&\left. -\sqrt{1-x_{{\rm B}}^{2}}-\frac{1-h_{0}x_{{\rm B}}}{\sqrt{
1-x_{{\rm B}}^{2}}}\right),\label{int_uext}
\end{align}
with $x_{\rm E}=x(r_{\rm E})=r_0/r_{\rm E}$. A similar expression holds for the second integral over $[r_{\rm B},r_{\rm S}]$, with $r_{\rm E}$ substituted by $r_{\rm S}$. For the last integral of the interior region in (\ref{def}), we make use of (\ref{integ1}) and (\ref{integ2}), as well as
\begin{equation}
\int \dfrac{x{\rm d}x}{\left( 1-x^{2}\right) \sqrt{1-x^{2}}\sqrt{x^{2}-\lambda _{0}^{2}}}= \frac{1}{1-\lambda _{0}^{2}}\frac{
\sqrt{x^{2}-\lambda _{0}^{2}}}{\sqrt{1-x^{2}}},  \label{integ4}
\end{equation}
to obtain
\begin{align}
 \int_{r_{0}}^{r_{{\rm B}}}\frac{{\rm d}r}{u_{{\rm int}}(r)}&\simeq\frac{\pi }{2
}-\arcsin x_{{\rm B}}+3 \frac{\delta _{{\rm B}}}{2}\frac{x_{{\rm B}}}{\sqrt{1-x_{{\rm B}}^{2}}}\nonumber\\
&\times\frac{ \sqrt{1-\lambda_{0}^{2}{}}-\sqrt{1-\lambda _{{\rm B}}^{2}} }{\lambda _{{\rm B}}^{2}}\sqrt{1-\lambda _{{\rm B}}^{2}}, \label{int_uint1}
\end{align}
or, equivalently, in terms of $h_0$,
\begin{align}
\int_{r_{0}}^{r_{{\rm B}}}\frac{{\rm d}r}{u_{{\rm int}}(r)} &\simeq
\frac{\pi }{2}-\arcsin x_{{\rm B}} + \frac{\delta _{0}}{2} \nonumber\\
&\times\left[ \frac{1-h{}_{0}x_{{\rm B}}}{\sqrt{1-x_{{\rm B}}^{2}}} -\left( 1 -2x_{{\rm B}}^{2}\right) \sqrt{1-x_{{\rm B}}^{2}}\right].\label{int_uint2}
\end{align}
It is straightforward to see that the two apparent singularities at $r=r_0$ in the integrals (\ref{integ2}) and (\ref{integ4}) cancel upon summation, thereby ensuring regularity. In fact, this could have been avoided outright using the identity in Eq.~(\ref{identity}), which allows the integral to be finite at $r=r_0$. Inserting the evaluated integrals back into the deflection formula~(\ref{def}), we obtain 
\begin{align}
-\varphi _{{\rm S}} &\simeq -\arcsin \frac{r_{0}}{r_{{\rm E}}}-\arcsin 
\frac{r_{0}}{r_{{\rm S}}}+\frac{GM}{r_{0}}\left[ \sqrt{1-\frac{r_{0}^{2}}{r_{{\rm E}}^{2}}}\right.\nonumber \\
&+\sqrt{1-\frac{r_{0}^{2}}{r_{{\rm S}}^{2}}}+\frac{1-h_{0}r_{0}/r_{{\rm E}}}{\sqrt{1-r_{0}^{2}/r_{{\rm E}}^{2}}}+\frac{1-h_{0}r_{0}/r_{{\rm S}}}{\sqrt{1-r_{0}^{2}/r_{{\rm S}}^{2}}} \nonumber \\
&-\left. 4\sqrt{1-\frac{r_{0}^{2}}{r_{{\rm B}}^{2}}}\left( 1-\frac{r_{0}^{2}}{r_{{\rm B}}^{2}}\right) \right],\label{varphi_S}
\end{align}
where $h_{0}$ is approximated to leading order in $\lambda_{\rm B}$ by
\begin{equation}
 h_{0}\simeq \frac{1}{2}\dfrac{r_{0}}{r_{{\rm B}}}\left[ 3-\dfrac{r_{0}^{2}}{
r_{{\rm B}}^{2}} +\frac{3}{4}\left( 1-\frac{r_{0}^{2}}{r_{{\rm B
}}^{2}}\right) ^{2}\lambda _{{\rm B}}^{2}\right]. \label{h} 
\end{equation}

To obtain the expression of $-\varphi _{{\rm S}}^\prime$ from the lower trajectory, it suffices to take a negative $J$, which corresponds to replacing $r_0$ by $r_{0}^\prime$. We then get
\begin{align}
-\varphi _{{\rm S}}^\prime &\simeq \arcsin \frac{r_{0}^\prime}{r_{{\rm E}}}+\arcsin 
\frac{r_{0}^\prime}{r_{{\rm S}}}-\frac{GM}{r_{0}^\prime}\left[ \sqrt{1-\frac{r_{0}^{\prime2}}{r_{{\rm E}}^{2}}}\right.\nonumber \\
&+\sqrt{1-\frac{r_{0}^{\prime2}}{r_{{\rm S}}^{2}}}+\frac{1-h_{0}^\prime r_{0}^\prime/r_{{\rm E}}}{\sqrt{1-r_{0}^{\prime2}/r_{{\rm E}}^{2}}}+\frac{1-h_{0}^\prime r_{0}^\prime/r_{{\rm S}}}{\sqrt{1-r_{0}^{\prime2}/r_{{\rm S}}^{2}}} \nonumber \\
&-\left. 4\sqrt{1-\frac{r_{0}^{\prime2}}{r_{{\rm B}}^{2}}}\left( 1-\frac{r_{0}^{\prime2}}{r_{{\rm B}}^{2}}\right) \right],\label{varphi_Sprime}
\end{align}
where $h^\prime=h(r_{0}^\prime)$. 

Because the Earth (and also the source) is located at a distance much larger than the characteristic lens scale ($r_{\rm B}\ll r_{\rm E} $), we can safely adopt the small-angle approximations, $\alpha\ll1$ and $\alpha^\prime\ll1$, with $2GM/r_{{\rm E}}\ll \lambda_{{\rm E}}^2 (=\Lambda r_{{\rm E}}^2/3)$. This leads to (\ref{r0rel})
\begin{equation}
r_{0}\simeq\frac{\alpha r_{\text{E}}}{\sqrt{1-\lambda _{\text{E}}^{2}}},\quad r_{0}^{\prime}\simeq \frac{\alpha ^{\prime }r_{\text{E}}}{\sqrt{1-\lambda_{\text{E}}^{2}}}.\label{r0approx}
\end{equation}
Using this, we can then further simplify (\ref{varphi_S}) and (\ref{varphi_Sprime}) as follows,
\begin{align}
-\varphi_{\text{S}}&=\frac{4GM\sqrt{1-\lambda_{\text{E}}^{2}}}{\alpha r_{\text{E}}}\left[1-\left(1-\dfrac{\alpha^{2}}{\beta ^{2}}\right) ^{\frac{3}{2}}\right]-\alpha\frac{1+r_{\text{E}}/r_{\text{S}}}{\sqrt{1-\lambda_{\text{E}}^{2}}},\\
-\varphi_{\text{S}}^{\prime}&=\alpha^{\prime}\frac{1+r_{\text{E}}/r_{\text{S}}}{\sqrt{1-\lambda_{\text{E}}^{2}}}-\frac{4GM\sqrt{1-\lambda_{\text{E}}^{2}}}{\alpha^{\prime}r_{\text{E}}}\left[1-\left(1-\dfrac{\alpha^{\prime 2}}{\beta ^{2}}\right)^{\frac{3}{2}}
\right] ,
\end{align}
where we have introduced
\begin{equation}
\beta\simeq \frac{r_{\text{B}}}{r_{\text{E}}}\sqrt{1-\lambda _{\text{E}}^{2}}\label{beta},
\end{equation}
related to the boundary radius $r_{\rm B}$ by analogy with (\ref{r0approx}). Given that both photons start from the same point, the source, $\varphi_{\text{S}}=\varphi_{\text{S}}^{\prime}$. Hence, this leads to an expression for the mass of the lens,
\begin{align}
M&=\frac{\alpha\alpha^{\prime}r_{\text{E}}}{4G}\frac{1+r_{\text{E}}/r_{\text{S}}}{1-\lambda_{\text{E}}^{2}}\nonumber\\
&\times\left(1-\frac{\left(1-\alpha^{2}/\beta^{2}\right)^{\frac{3}{2}}}{1+\alpha/\alpha^{\prime}}-\frac{\left(1-\alpha^{\prime2}/\beta^{2}\right)^{\frac{3}{2}}}{1+\alpha^{\prime }/\alpha}\right)^{-1}\label{mass},
\end{align}
and
\begin{align}
-\varphi_{\text{S}}&=-\varphi_{\text{S}}^{\prime}\simeq\frac{1+r_{\text{E}}/r_{\text{S}}}{\sqrt{1-\lambda_{\text{E}}^{2}}}\left\{\alpha^{\prime}\left[1-\left(1-\dfrac{\alpha^{2}}{\beta ^{2}}\right)^{\frac{3}{2}}\right]^{\phantom{-\frac{A^2}{A}}}\right.\\
&\left.\times\left(1-\frac{\left(1-\alpha^{2}/\beta^{2}\right)^{\frac{3}{2}}}{1+\alpha/\alpha^{\prime}}-\frac{\left(1-\alpha^{\prime2}/\beta^{2}\right)^{\frac{3}{2}}}{1+\alpha^{\prime }/\alpha}\right)^{-1}-\alpha\right\}.
\end{align}
As expected, the case of perfect alignment of the source with both the lens and the Earth, corresponding to
$\varphi_{\text{S}}=0$, is verified for $\alpha=\alpha^{\prime}$ due to symmetry. 
\subsection{Time delay}
The objective now is to compute the proper time delay of the photon traveling along the upper trajectory relative to that along the lower trajectory, $\Delta\tau=\sqrt{1-\lambda_{\rm E}^2}\Delta t$ with $\Delta t=t_{\rm S}^\prime-t_{\rm S}$.

Let us first determine the time elapsed during the upper photon's propagation from $r_{\rm B}$ to $r_{\rm E}$ and  from $r_{0}$ to $r_{\rm B}$  (\ref{Ttravel}). Performing a first-order expansion in $\delta _{0}$ yields
\begin{align}
v_{\text{ex}}(x)^{-1} &\simeq \frac{x^{2}\sqrt{f_{0}}}{(x^{2}-\lambda
_{0}^{2})\sqrt{1-x^{2}}}\nonumber\\
&\times\left[ 1+\frac{\delta _{0}}{2}\left( \frac{
h{}_{0}-x^{3}}{1-x^{2}}+2\frac{x^{3}}{x^{2}-\lambda _{0}^{2}{}}\right)
\right] \label{vex}, \\
v_{\text{in}}(x)^{-1} &\simeq \frac{x^{2}\sqrt{f_{0}}}{(x^{2}-\lambda
_{0}^{2})\sqrt{1-x^{2}}}\nonumber\\
&\times\left[ 1+\frac{\delta _{0}}{2}\left( \frac{
h_{0}-h(x)x^{3}}{1-x^{2}}+\frac{h(x)x^{3}+x_{\text{B}}^{3}}{x^{2}-\lambda
_{0}^{2}}\right) \right]\label{vin},
\end{align}
with
\begin{equation}
h(x)x^{3}=x_{\text{B}}^{3}\left( 3\frac{x-\sqrt{1-\lambda _{\text{B}}^{2}}
\sqrt{x^{2}-\lambda _{0}^{2}}}{\lambda _{0}^{2}}x-2 \right).
\end{equation}
It follows that, to integrate Eq.~(\ref{vex}), one must employ the following integrals:
\begin{alignat}{2}
&\int \frac{\text{d}x}{(x^{2}-\lambda _{0}^{2})\sqrt{1-x^{2}}}&&=-\mu (x), \label{mu}\\
&\int \frac{\text{d}x}{(x^{2}-\lambda _{0}^{2})( 1-x^{2}) ^{\frac{3}{2}}}&&=\frac{1}{1-\lambda _{0}^{2}}\left( \frac{x}{\sqrt{1-x^{2}}}-\mu (x)\right),\\
&\int \frac{x^{3}\text{d}x}{(x^{2}-\lambda _{0}^{2})( 1-x^{2})^{\frac{3}{2}}}&&=\frac{1}{1-\lambda _{0}^{2}}\left( \frac{1}{\sqrt{1-x^{2}}}-\lambda _{0}^{2}\nu (x)\right), \\
&\int \frac{x^{3}\text{d}x}{(x^{2}-\lambda _{0}^{2})^{2}\sqrt{1-x^{2}}}&&=\frac{-1}{2(1-\lambda _{0}^{2})}\nonumber\\
&&&\times\left( \lambda _{0}^{2}\frac{\sqrt{1-x^{2}}}{x^{2}-\lambda _{0}^{2}}+(2-\lambda _{0}^{2})\nu (x)\right),
\end{alignat}
where $\mu(x)$ and $\nu(x)$ are defined by
\begin{align}
\mu (x) &=\frac{1}{\lambda _{0}\sqrt{1-\lambda _{0}^{2}}}{\rm arctanh}
\frac{\lambda(x)\sqrt{1-x^{2}}}{\sqrt{1-\lambda _{0}^{2}}}, \\
\nu (x) &=\frac{1}{\sqrt{1-\lambda _{0}^{2}}}{\rm arctanh}\frac{\sqrt{
1-x^{2}}}{\sqrt{1-\lambda _{0}^{2}}}.
\end{align}
Note that we omit an additive imaginary constant in $\mu(x)$, as it cancels upon evaluation. The result is
\begin{align}
\int_{r_{\text{B}}}^{r_{\text{E}}}\frac{\text{d}r}{v_{\text{ex}}(r)}&\simeq r_{0}
\sqrt{f_{0}}\left[ \mu _{\text{E}}-\mu _{\text{B}}+\frac{\delta _{0}}{2}
\left( \kappa _{\text{E}}-\kappa_{\text{B}}\phantom{\frac{A}{A}}\right.\right.\nonumber\\
&\left.\left.+h{}_{0}\frac{\mu _{\text{E}}-\mu _{\text{B}}-\varepsilon_{\text{E}}+\varepsilon _{\text{B}}}{1-\lambda_{0}^{2}}+2\nu_{\text{E}}-2\nu _{\text{B}}\right) \right]\label{rBrE},
\end{align}
where $\mu _{\text{E}}=\mu (x_{\text{E}})$, $\mu _{\text{B}}=\mu (x_{\text{B}})$, $\nu_{\text{E}} =\nu (x_{\text{E}})$, $\nu _{\text{B}}=\nu (x_{\text{B}})$, $\kappa _{%
\text{E}}=\kappa (x_{\text{E}})$, and $\kappa _{\text{B}}=\kappa (x_{\text{B}})$ are given by the functions
\begin{equation}
\kappa (x)=\frac{1}{[1-\lambda (x)^{2}]\sqrt{1-x^{2}}},\quad
\varepsilon (x)=\frac{x}{\sqrt{1-x^{2}}}.
\end{equation}
As for Eq.\~(\ref{vin}), we make use, in addition to (\ref{mu}), of the following integrals: 
\begin{alignat}{2}
&\int \dfrac{x^{2}\text{d}x}{(x^{2}-\lambda _{0}^{2})( 1-x^{2})
^{\frac{3}{2}}}&&=\frac{1}{1-\lambda _{0}^{2}}\left( \frac{x}{\sqrt{1-x^{2}}}
-\lambda _{0}^{2}\mu (x)\right),  \\
&\int \dfrac{x\text{d}x}{\sqrt{x^{2}-\lambda _{0}^{2}}\left( 1-x^{2}\right)
^{\frac{3}{2}}}&&=\frac{1}{1-\lambda _{0}^{2}}\frac{\sqrt{x^{2}-\lambda
_{0}^{2}}}{\sqrt{1-x^{2}}}, \\
&\int \dfrac{x^{2}\text{d}x}{(x^{2}-\lambda _{0}^{2})^{2}\sqrt{1-x^{2}}}&&=
\frac{-1}{2(1-\lambda _{0}^{2})}\left( \frac{x\sqrt{1-x^{2}}}{x^{2}-\lambda_{0}^{2}}+\mu (x)\right) , \\
&\int \dfrac{x\text{d}x}{(x^{2}-\lambda _{0}^{2})^{\frac{3}{2}}\sqrt{1-x^{2}
}}&&=\frac{-1}{1-\lambda _{0}^{2}}\frac{\sqrt{1-x^{2}}}{\sqrt{x^{2}-\lambda
_{0}^{2}}}, \\
&\int \dfrac{\text{d}x}{(x^{2}-\lambda _{0}^{2})^{2}\sqrt{1-x^{2}}}&&=\frac{-1}{2\lambda _{0}^{2}(1-\lambda _{0}^{2})}\nonumber\\
&&&\times\left( \frac{x\sqrt{1-x^{2}}}{x^{2}-\lambda _{0}^{2}}-(1-2\lambda _{0}^{2})\mu (x)\right).
\end{alignat}
We obtain
\begin{align}
\int_{r_{0}}^{r_{\text{B}}}\frac{\text{d}r}{v_{\text{in}}(x)}&\simeq r_{0}\sqrt{
f_{0}}\left[ \mu_{\text{B}}+\frac{\delta_{0}}{2}\left(\kappa_{\text{B}}+h_{0}\frac{\mu_{\text{B}}-\varepsilon_{\text{B}}}{1-\lambda_{0}^{2}}\phantom{\frac{\sqrt{A^2}}{\sqrt{A_{0}^2}}}\right.\right.\nonumber\\
&\left.\left.+2\frac{x_{\text{B}}\mu_{\text{B}}-\sqrt{1-x_{\text{B}}^{2}}}{%
\lambda _{\text{B}}^{2}}\right) \right].
\end{align}
Similarly, the propagation time  from $r_{\rm B}$ to $r_{\rm S}$ is obtained by replacing $r_{\rm E}$ with $r_{\rm S}$ in  (\ref{rBrE}). Thus, the travel time of the upper photon (\ref{Ttravel}) is given by
\begin{equation}
-t_{\text{S}}\simeq r_{0}\sqrt{f_{0}}\left( \mu_{\text{E}}+\mu_{\text{S}}+\frac{
\delta _{0}}{2}\frac{I}{\sqrt{f_{0}}} \right),
\end{equation}
where, for simplicity, we have introduced the function $I$,
\begin{align}
\frac{I}{\sqrt{f_{0}}}&= \kappa_{\text{E}}+\kappa_{\text{S}}+h_{0}\frac{\mu_{\text{E}}+\mu_{\text{S}}-\varepsilon_{\text{E}}-\varepsilon_{\text{S}}}{1-\lambda _{0}^{2}}\nonumber\\
&+4\frac{x_{\text{B}}\mu_{\text{B}}-\sqrt{1-x_{\text{B}}^{2}}}{\lambda _{\text{B}}^{2}}+2\nu_{\text{E}}+2\nu_{\text{S}}-4\nu_{\text{B}},
\end{align}
using the definitions $\mu_{\text{S}}=\mu (x_{\text{S}})$, $\nu_{\text{S}}=\nu (x_{\text{S}})$, $\kappa_{\text{S}}=\kappa (x_{\text{S}})$, and $\varepsilon_{\text{S}}=\varepsilon (x_{\text{S}})$. Consequently, the time delay
$\Delta t=t_{\text{S}}^{\prime}-t_{\text{S}}$ is
\begin{equation}
 \Delta t\simeq r_{0}\sqrt{f_{0}}( \mu_{\text{E}}+\mu_{\text{S}})-r_{0}^{\prime}\sqrt{f_{0}^{\prime}}(\mu_{\text{E}}^{\prime}+\mu_{\text{S}}^{\prime})+GM\Delta I\label{totalTD},
\end{equation}
where $\Delta I=\Delta I_{\rm E} +\Delta I_{\rm S}+\Delta I_{\rm B}$, with
\begin{align}
\Delta I_{\rm E}&=\sqrt{f_{0}}\left(\kappa_{\text{E}}+h_{0}\frac{\mu_{\text{E}}-\varepsilon_{\text{E}}}{1-\lambda _{0}^{2}}+2\nu_{\text{E}}\right)\nonumber\\
&-\sqrt{f_{0}^{\prime}}\left(\kappa_{\text{E}}^{\prime}+h_{0}^{\prime}\frac{\mu_{\text{E}}^{\prime}-\varepsilon_{\text{E}}^{\prime}}{1-\lambda_{0}^{{\prime}2}}+2\nu_{\text{E}}^{\prime}\right),\label{DeltaE}\\
\Delta I_{\rm B}&=4\sqrt{f_{0}}\left(\frac{x_{\text{B}}\mu_{\text{B}}-\sqrt{1-x_{\text{B}}^{2}}}{\lambda_{\text{B}}^{2}}-\nu_{\text{B}}\right)\nonumber\\
&-4\sqrt{f_{0}^{\prime}}\left(\frac{x_{\text{B}}^{\prime}\mu_{\text{B}}^{\prime}-\sqrt{1-x_{\text{B}}^{{\prime}2}}}{\lambda _{\text{B}}^{{\prime}2}}-4\nu_{\text{B}}^{\prime}\right).
\end{align}
All the primed quantities refer to the lower photon, obtained by replacing $r_{0}$ with $r_{0}^\prime$. The expression of $\Delta I_{\rm S}$ is similar to (\ref{DeltaE}), substituting $r_{\rm E}$ by $r_{\rm S}$. After a lengthy calculation, the differences in (\ref{totalTD}) are evaluated at leading order in $\delta_{0}$, $x_{\rm E}$, $x_{\rm S}$, and $\lambda_{0}$ as follows: 
\begin{align}
r_{0}\sqrt{f_{0}}\mu_{\text{E}}-r_{0}^{\prime}\sqrt{f_{0}^{\prime}}\mu_{\text{E}}^{\prime }&\simeq \frac{r_{0}}{2}\left( \frac{r_{0}^{\prime 2}}{r_{0}^{2}}-1\right) x_{\text{E}}-GM\nonumber\\
&\times\left[h_{0}\left( 1-\frac{r_{0}}{r_{0}^{\prime }}\frac{h_{0}^{\prime }}{h{}_{0}}\right)\frac{{\rm arctanh}\lambda _{\text{E}}}{\lambda_{0}} \right.\nonumber\\
&+\left.\frac{\delta _{0}}{4}h_{0}^{2}\left( 1-\frac{r_{0}^{2}}{r_{0}^{\prime 2}}\frac{h_{0}^{\prime 2}}{h_{0}^{2}}\right) \frac{{\rm arctanh}\lambda _{\text{E}}}{\lambda _{0}}\right],
\end{align}
and
\begin{align}
\Delta I_{\text{E}}&\simeq h_{0}\left( 1 -\frac{h_{0}^{\prime } }{h_{0}}\frac{r_{0}}{r_{0}^{\prime }}\right)\frac{{\rm arctanh}\lambda _{\text{E}}}{\lambda_{0}}+2\ln\frac{r_{0}^{\prime }}{r_{0}}\nonumber\\
&-\frac{\delta _{0}}{2}h_{0}^{2} \left(1- \frac{r_{0}^{2}}{r_{0}^{\prime2}}\frac{h{}_{0}^{\prime 2}}{h_{0}^{2}}\right)\frac{{\rm arctanh}\lambda _{\text{E}}}{\lambda _{0}},\\
\Delta I_{\text{B}}&\simeq \frac{4}{3}\left( (1+2x_{\text{B}}^{2})
\sqrt{1-x_{\text{B}}^{2}}-(1+2x_{\text{B}}^{\prime 2})\sqrt{1-x_{
\text{B}}^{\prime 2}}\right)\nonumber\\
&-4\left( {\rm arctanh}\sqrt{1-x_{\text{B}}^{2}}
-{\rm arctanh}\sqrt{1-x_{\text{B}}^{\prime 2}}\right).
\end{align}
In these calculations, we find $\sqrt{f_{0}}\kappa_{\text{E}}-\sqrt{f_{0}^{\prime }}\kappa_{\text{E}}^{\prime }\simeq \mathcal{O}(x_{\text{E}}^{2})$, and $\sqrt{f_{0}}h_{0}\varepsilon_{\text{E}}/(1-\lambda_{0}^{2})-\sqrt{f_{0}^{\prime }}h_{0}^{\prime}\varepsilon_{\text{E}}^{\prime}/(1-\lambda_{0}^{\prime 2})\simeq \mathcal{O}(x_{\text{E}})$, so these terms are negligible when multiplied by $GM$. Isolating the interior correction relative to the exterior SdS time delay $\Delta t_{\text{ex}}$ \cite{Schu2}, the total time delay can be written as
\begin{equation}
\Delta t\simeq \Delta t_{\text{ex}}+4GM\mathcal{E}_{\text{in}},
\end{equation}
where
\begin{align}
\Delta t_{\text{ex}}&=\frac{r_{0}}{2}\left(\frac{r_{0}^{\prime2}}{r_{0}^{2}}-1\right)\left(\frac{r_{0}}{r_{\text{E}}}+\frac{r_{0}}{r_{\text{S}}}\right)+4GM\left[\ln \frac{r_{0}^{\prime }}{r_{0}}\right.\nonumber\\
&\left. -\frac{3\delta _{0}}{16}\left(1-\frac{r_{0}^{2}}{r_{0}^{\prime 2}}\right)\frac{{\rm arctanh}\lambda _{\text{E}}+{\rm arctanh}\lambda _{\text{S}}}{\lambda _{0}}\right],
\end{align}
and
\begin{align}
\mathcal{E}_{\text{in}}&=\frac{1}{3}\left(1+2\frac{r_{0}^{2}}{r_{\text{B}}^2}\right)\sqrt{1-\frac{r_{0}^{2}}{r_{\text{B}}^2}}-\frac{1}{3}\left(1+2\frac{r_{0}^{\prime2}}{r_{\text{B}}^2}\right)\sqrt{1-\frac{r_{0}^{\prime2}}{r_{\text{B}}^2}} \nonumber\\
 &-{\rm arctanh}\sqrt{1-\frac{r_{0}^{2}}{r_{\text{B}}^2}}+{\rm arctanh}\sqrt{1-\frac{r_{0}^{\prime2}}{r_{\text{B}}^2}}+\frac{3\delta _{0}}{16}(1-h_{0}^{2})\nonumber\\
&\times\left(1-\frac{1-h_{0}^{\prime 2}}{1-h_{0}^{2}}\frac{r_{0}^{2}}{r_{0}^{\prime2}}\right)\frac{{\rm arctanh}\lambda_{\text{E}}+{\rm arctanh}\lambda_{\text{S}}}{\lambda _{0}}.
\end{align}
Finally, we express the proper time delay in terms of measurable quantities as
\begin{equation}
\Delta \tau\simeq \Delta \tau_{\text{ex}}+4GM\sqrt{1-\frac{\Lambda}{3}r_{\rm E}^2}\mathcal{E}_{\text{in}},
\end{equation}
using (\ref{r0approx}) and (\ref{beta}), where $\Delta \tau_{\text{ex}}=\sqrt{1-\Lambda r_{\text{E}}^{2}/3}\Delta t_{\text{ex}}$\cite{Schu2}, with
\begin{align}
\Delta t_{\text{ex}}&\simeq\frac{\alpha^2r_{\text{E}}}{2}\left(\frac{\alpha^{\prime2}}{\alpha^{2}}-1\right)\frac{1+r_{\text{E}}/r_{\text{S}}}{\sqrt{1-\Lambda r_{\text{E}}^{2}/3}}+4GM\nonumber\\
&\times\left[\ln \frac{r_{0}^{\prime }}{r_{0}} -\frac{3GM(1-\Lambda r_{\text{E}}^{2}/3)}{8\alpha^{2}r_{\text{E}}^{2}}\left(1-\frac{\alpha^{2}}{\alpha^{\prime2}}\right)^{\phantom{A}}\right.\nonumber\\
&\left.\times\frac{{\rm arctanh}\sqrt{\Lambda r_{\text{E}}^2/3 }+{\rm arctanh}\sqrt{\Lambda r_{\text{S}}^2/3 }}{\sqrt{\Lambda/3 }}\right],
\end{align}
and
\begin{align}
\mathcal{E}_{\text{in}}&=\frac{1}{3}\left(1+2\frac{\alpha^{2}}{\beta^2}\right)\sqrt{1-\frac{\alpha^{2}}{\beta^2}}-\frac{1}{3}\left(1+2\frac{\alpha^{\prime2}}{\beta^2}\right)\sqrt{1-\frac{\alpha^{\prime2}}{\beta^2}} \nonumber\\
 &-{\rm arctanh}\sqrt{1-\frac{\alpha^{2}}{\beta^2}}+{\rm arctanh}\sqrt{1-\frac{\alpha^{\prime2}}{\beta^2}}\nonumber\\
&+\frac{3GM(1-\Lambda r_{\text{E}}^{2}/3)}{8\alpha^{2}r_{\text{E}}^{2}}(1-h_{0}^{2})\left(1-\frac{1-h_{0}^{\prime 2}}{1-h_{0}^{2}}\frac{\alpha^{2}}{\alpha^{\prime2}}\right)\nonumber\\
&\frac{{\rm arctanh}\sqrt{\Lambda r_{\text{E}}^2/3 }+{\rm arctanh}\sqrt{\Lambda r_{\text{S}}^2/3 }}{\sqrt{\Lambda/3 }},
\end{align}
where the mass $M$ is given by (\ref{mass}), $h_0$ and $h_{0}^\prime$ (\ref{h}) are approximated by
\begin{equation}
 h_{0}\simeq \frac{1}{2}\frac{\alpha}{\beta}\left( 3-\dfrac{\alpha^{2}}{\beta^{2}}\right),\quad  h_{0}^\prime\simeq \frac{1}{2}\frac{\alpha^\prime}{\beta}\left( 3-\dfrac{\alpha^{\prime2}}{\beta^{2}}\right). 
\end{equation}
Again,  in the aligned configuration  ($\varphi_{\rm S}=0$), one can straightforwardly verify that the time delay vanishes ($\Delta\tau=0$)  for symmetry reason when $\alpha=\alpha^{\prime}$.
\section{\label{IV}Numerical Results and Discussion}
In this section, we present numerical evaluations of the gravitational lensing time delays for two representative astrophysical lenses: a galaxy-scale lens with boundary radius $r_{\rm B}=25~\mathrm{kpc}$, and a cluster-scale lens with $r_{\rm B}=250~\mathrm{kpc}$. In both cases, the source and observer are placed at equal radial distances $r_{\rm S}=r_{\rm E}=3~\mathrm{Gpc}$, and the cosmological constant is fixed at $\Lambda=10^{-52}~\mathrm{m^{-2}}$. We compute the total time delay $\Delta \tau$ and the exterior-only time delay $\Delta \tau_{\text{ex}}$ as functions of the reception angle $\alpha$.
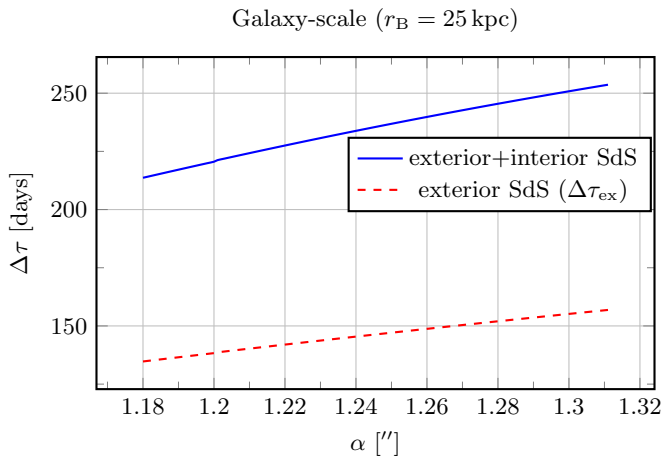
\begin{figure}[htbp]
    \centering
    \begin{tikzpicture}
\begin{axis}[
    width=9cm,
    height=6cm,
    xlabel={$\alpha$ [$^{\prime\prime}$]},
    ylabel={$\Delta \tau$ [days]},
    title={Galaxy-scale ($r_{\rm B}=25\,\mathrm{kpc}$)},
    grid=major,
  thick,
    minor tick num=1,
    legend style={at={(0.45,0.76)},anchor=north west},
]

\addplot[smooth, thick, blue] coordinates {
(1.180, 213.656) (1.182, 214.356) (1.184, 215.054) (1.186, 215.751) (1.188, 216.445) (1.190, 217.137) (1.192, 217.828) (1.194, 218.515) (1.196, 219.199) (1.198, 219.881) (1.200, 220.560) (1.201, 221.237) (1.203, 221.911) (1.205, 222.582) (1.207, 223.250) (1.209, 223.914) (1.211, 224.576) (1.213, 225.235) (1.215, 225.891) (1.217, 226.543) (1.219, 227.193) (1.221, 227.839) (1.223, 228.482) (1.225, 229.123) (1.227, 229.760) (1.229, 230.393) (1.231, 231.024) (1.233, 231.651) (1.235, 232.275) (1.237, 232.896) (1.239, 233.513) (1.241, 234.127) (1.243, 234.737) (1.245, 235.345) (1.247, 235.949) (1.249, 236.550) (1.251, 237.148) (1.253, 237.742) (1.255, 238.334) (1.257, 238.922) (1.259, 239.508) (1.261, 240.090) (1.263, 240.669) (1.265, 241.244) (1.267, 241.817) (1.269, 242.386) (1.271, 242.952) (1.273, 243.515) (1.275, 244.075) (1.277, 244.632) (1.279, 245.186) (1.281, 245.737) (1.283, 246.284) (1.285, 246.829) (1.287, 247.371) (1.289, 247.910) (1.291, 248.446) (1.293, 248.979) (1.295, 249.509) (1.297, 250.036) (1.299, 250.560) (1.301, 251.081) (1.303, 251.599) (1.305, 252.114) (1.307, 252.626) (1.309, 253.135) (1.311, 253.641)
};
\addlegendentry{exterior+interior SdS}

\addplot[dashed, thick, red] coordinates {
(1.180, 134.735) (1.182, 135.097) (1.184, 135.458) (1.186, 135.818) (1.188, 136.177) (1.190, 136.536) (1.192, 136.894) (1.194, 137.251) (1.196, 137.607) (1.198, 137.962) (1.200, 138.316) (1.201, 138.670) (1.203, 139.023) (1.205, 139.375) (1.207, 139.726) (1.209, 140.076) (1.211, 140.426) (1.213, 140.775) (1.215, 141.123) (1.217, 141.471) (1.219, 141.817) (1.221, 142.164) (1.223, 142.509) (1.225, 142.854) (1.227, 143.198) (1.229, 143.541) (1.231, 143.883) (1.233, 144.225) (1.235, 144.565) (1.237, 144.905) (1.239, 145.244) (1.241, 145.582) (1.243, 145.920) (1.245, 146.256) (1.247, 146.591) (1.249, 146.926) (1.251, 147.260) (1.253, 147.593) (1.255, 147.925) (1.257, 148.255) (1.259, 148.586) (1.261, 148.915) (1.263, 149.243) (1.265, 149.571) (1.267, 149.897) (1.269, 150.223) (1.271, 150.548) (1.273, 150.872) (1.275, 151.196) (1.277, 151.518) (1.279, 151.840) (1.281, 152.160) (1.283, 152.480) (1.285, 152.799) (1.287, 153.117) (1.289, 153.435) (1.291, 153.751) (1.293, 154.067) (1.295, 154.382) (1.297, 154.696) (1.299, 155.010) (1.301, 155.322) (1.303, 155.634) (1.305, 155.945) (1.307, 156.255) (1.309, 156.565) (1.311, 156.874)
};
\addlegendentry{exterior SdS ($\Delta\tau_{\rm ex}$)}

\end{axis}
\end{tikzpicture}
   
\caption{Evolution of time delay versus reception angle $\alpha\in[1.18^{\prime\prime},1.31^{\prime\prime}]$ with $\alpha^\prime=\alpha/0.9$ for a galaxy lens in matched Kottler spacetime.}
    \label{TD_galaxy}
\end{figure}
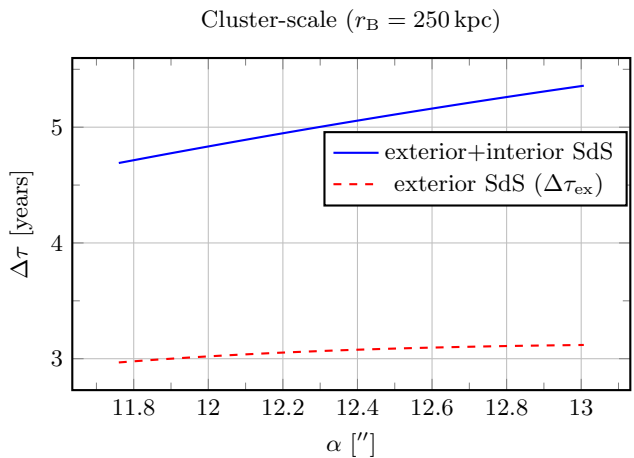
\begin{figure}[htbp]
    \centering
    \begin{tikzpicture}
\begin{axis}[
    width=9cm,
    height=6cm,
    xlabel={$\alpha$ [$^{\prime\prime}$]},
    ylabel={$\Delta \tau$ [years]},
    title={Cluster-scale ($r_{\rm B}=250\,\mathrm{kpc}$)},
    grid=major,
  thick,
    minor tick num=1,
    legend style={at={(0.45,0.78)},anchor=north west},
]

\addplot[thick, smooth,blue] coordinates {
(11.760, 1713.228/365.25) (11.803, 1722.774/365.25) (11.846, 1732.238/365.25) (11.889, 1741.620/365.25) (11.932, 1750.920/365.25) (11.975, 1760.139/365.25) (12.018, 1769.276/365.25) (12.061, 1778.333/365.25) (12.104, 1787.308/365.25) (12.147, 1796.203/365.25) (12.190, 1805.017/365.25) (12.233, 1813.749/365.25) (12.276, 1822.401/365.25) (12.319, 1830.972/365.25) (12.362, 1839.462/365.25) (12.405, 1847.872/365.25) (12.448, 1856.200/365.25) (12.491, 1864.448/365.25) (12.534, 1872.616/365.25) (12.577, 1880.702/365.25) (12.620, 1888.709/365.25) (12.663, 1896.635/365.25) (12.706, 1904.481/365.25) (12.749, 1912.246/365.25) (12.792, 1919.931/365.25) (12.835, 1927.535/365.25) (12.878, 1935.059/365.25) (12.921, 1942.503/365.25) (12.964, 1949.867/365.25) (13.007, 1957.150/365.25)
};
\addlegendentry{exterior+interior SdS}

\addplot[thick, smooth, dashed,red] coordinates {
(11.760, 1083.928/365.25) (11.803, 1087.774/365.25) (11.846, 1091.432/365.25) (11.889, 1094.905/365.25) (11.932, 1098.199/365.25) (11.975, 1101.319/365.25) (12.018, 1104.271/365.25) (12.061, 1107.061/365.25) (12.104, 1109.695/365.25) (12.147, 1112.179/365.25) (12.190, 1114.521/365.25) (12.233, 1116.727/365.25) (12.276, 1118.803/365.25) (12.319, 1120.756/365.25) (12.362, 1122.590/365.25) (12.405, 1124.312/365.25) (12.448, 1125.926/365.25) (12.491, 1127.438/365.25) (12.534, 1128.852/365.25) (12.577, 1130.172/365.25) (12.620, 1131.403/365.25) (12.663, 1132.550/365.25) (12.706, 1133.616/365.25) (12.749, 1134.605/365.25) (12.792, 1135.519/365.25) (12.835, 1136.362/365.25) (12.878, 1137.139/365.25) (12.921, 1137.851/365.25) (12.964, 1138.501/365.25) (13.007, 1139.093/365.25)
};
\addlegendentry{exterior SdS ($\Delta\tau_{\rm ex}$)}

\end{axis}
\end{tikzpicture}
   
\caption{Evolution of time delay versus reception angle $\alpha\in[11.76^{\prime\prime},13.07^{\prime\prime}]$ with $\alpha^\prime=\alpha/0.9$ for a cluster lens in matched Kottler spacetime.}
    \label{TD_cluster}
\end{figure}
The analysis strictly respects the constraint $\alpha=0.9\alpha^\prime$, with $\alpha^\prime\in[0.9\beta,\beta]$, ensuring the validity of applying the geodesic integration to first order in the ratio of the Schwarzschild radius to the pericenters, $r_0$ and $r_{0}^\prime$ or equivalently to the fluid radius $r_{\rm B}$. For the galaxy-scale lens, the mass spans approximately $1.7\times10^{11}M_{\odot }\lesssim M\lesssim1.9\times10^{11}M_{\odot }$, reflecting typical galactic mass scales, while for the cluster-scale lens, the mass corresponds to $1.7\times10^{14}M_{\odot }\lesssim M\lesssim1.9\times10^{14}M_{\odot }$. The corresponding plots of the functions $\Delta \tau(\alpha)$ and $\Delta \tau_{\text{ex}}(\alpha)$ are illustrated in Figs.~\ref{TD_galaxy} and \ref{TD_cluster}.

For the galaxy-scale lens with $r_{\rm B}=25~{\rm kpc}$ and $M\sim(1.7-1.9)\times10^{11} M_\odot$, Fig.~\ref{TD_galaxy} shows that across $\alpha\in[1.18^{\prime\prime},1.31^{\prime\prime}]$ (with $\alpha^\prime=\alpha/0.9$), the total time delay lies in the range $\Delta\tau\sim214$--$254$~days, systematically exceeding the exterior-only SdS prediction $\Delta\tau_{\rm ex}\sim135$--$157$~days. The excess due to the interior contribution, $\Delta\tau-\Delta\tau_{\rm ex}\sim79$--$97$~days, is clearly visible and reflects the cumulative effect of the distributed mass inside the lens.

For the cluster-scale lens with $r_{\rm B}=250~{\rm kpc}$ and $M\sim(1.7-1.9)\times10^{14} M_\odot$, Fig.~\ref{TD_cluster} shows significantly larger delays: $\Delta\tau\simeq4.7$--$5.4$~years for $\alpha\in[11.8^{\prime\prime},13.1^{\prime\prime}]$ (with $\alpha^\prime=\alpha/0.9$), compared with $\Delta\tau_{\rm ex}\sim3.0$--$3.1$~years. The interior contribution amounts to $\sim1.7$--$2.3$~years, showing that the enhancement can reach nearly two years for cluster lenses.

\section{\label{V}Conclusion}
The analysis presented here is based on a fully relativistic treatment of light propagation in a matched Schwarzschild–de Sitter geometry. By integrating null geodesics for photon trajectories crossing both the interior and exterior regions, we obtained the source inclination angle $-\varphi_{\rm S}$ \cite{Guen5}, related to the total deflection angle, as well as the proper time delay between two distinct photon paths. In parallel, analytic approximations were derived for both observables, allowing us to isolate and quantify the contributions arising from the interior mass distribution relative to the vacuum Kottler exterior.

Numerical evaluation quantifies the corrections induced by the interior mass distribution relative to the exterior-only case. The matched SdS time delay always exceeds the exterior-only prediction by a substantial fraction, typically $\sim60$--$70\%$ across both scales. For galaxy-scale lenses, with $r_{\rm B}=25~{\rm kpc}$ and $M \sim 10^{11}M_\odot$, the interior correction contributes on the order of several tens of days, while for cluster-scale lenses, with $r_{\rm B}=250~{\rm kpc}$ and $M \sim 10^{14}M_\odot$, it reaches the level of several months to nearly two years. These results demonstrate that the interior contribution, though often neglected, plays a quantitatively significant role and must be incorporated for accurate modeling of strong lensing by extended astrophysical objects.

As a perspective, the present framework can be extended to dynamical cosmological backgrounds \cite{Schu1,Guen1,Guen2,Guen3,Guen4}, and to more realistic, non-uniform mass distributions \cite{Gabbanelli}. Such generalizations would allow for a direct comparison with observational lensing data in evolving universes and provide a more faithful modeling of galaxies and clusters beyond the idealized uniform-density case.

\bibliography{apssamp}

\end{document}